\documentclass[journal]{IEEEtran}
\usepackage[T1]{fontenc}
\usepackage{ifpdf}
\usepackage{cite}
\usepackage[pdftex]{graphicx}
\usepackage{amsmath}
\usepackage{amssymb}
\usepackage{bm}
\usepackage{array}
\usepackage{fixltx2e}
\usepackage{stfloats}

\usepackage{balance} 

\usepackage{mathtools} 

\usepackage{diagbox}

\usepackage{makecell}

\usepackage{threeparttable}

\usepackage{xcolor,soul}
\soulregister\cite7

\usepackage{algorithm}
\usepackage{algpseudocode}

\newtheorem{proposition}{Proposition}

\usepackage[hidelinks]{hyperref}

\begin{document}
\title{\vspace{-7mm}\LARGE Transmit and Receive Antenna Port Selection for Channel Capacity Maximization in Fluid-MIMO Systems}

\author{Christos N. Efrem and Ioannis Krikidis 
\vspace{-4mm} 
\thanks{This work was co-funded by the European Research Council (grant agreement no. 819819) and the Research \& Innovation Foundation of Cyprus (Larissa6G-Small Scale Infrastructures/1222/0087). The authors are with the Department of Electrical and Computer Engineering, University of Cyprus, Nicosia, Cyprus (e-mail: \{efrem.christos, krikidis\}@ucy.ac.cy). }  \vspace{-7mm} }

\markboth{T\MakeLowercase{his article has been accepted for publication in} IEEE W\MakeLowercase{ireless} C\MakeLowercase{ommunications} L\MakeLowercase{etters}, S\MakeLowercase{eptember} 2024. \copyright 2024 IEEE.}%
{}

\maketitle

\begin{abstract}

In this letter, we study a discrete optimization problem, namely, the maximization of channel capacity in fluid multiple-input multiple-output (fluid-MIMO) systems through the selection of antenna ports/positions at both the transmitter and the receiver. First, we present a new joint convex relaxation (JCR) problem by using an upper bound on the channel capacity and exploiting the binary nature of optimization variables. Then, we develop and analyze two optimization algorithms with different performance-complexity tradeoffs. The first algorithm is based on JCR and reduced exhaustive search (JCR\&RES), while the second on JCR and alternating optimization (JCR\&AO).  

\end{abstract}


\IEEEpeerreviewmaketitle

\vspace{-2mm}
\section{Introduction}

Multiple-input multiple-output (MIMO) technology has revolutionized wireless communications during the last decades, by achieving significant gains in terms of diversity and multiplexing. In classical MIMO, the positions of antennas are fixed and their separation distance should be large enough in order to fully exploit the multi-path propagation of signals (negligible spatial correlation) and avoid undesirable near-field phenomena (e.g., mutual coupling). Recently, a new technology has been proposed, namely, fluid antenna systems (FAS), where the antenna position can dynamically change and be adapted to the channel characteristics, thus providing more flexibility \cite{Wong2021}. In addition, FAS can enhance the system performance even when the available antenna positions are very close to each other, while mutual coupling completely disappears (as long as only one position/port is selected). 

Furthermore, the combination of MIMO with FAS is a promising technology and has already attracted the attention of the research community \cite{Wong2022}. Port selection strategies were proposed in classical FAS \cite{Wong2021} as well as in MIMO-FAS \cite{New2023} (i.e., FAS with activation of multiple ports at the transmitter and the receiver). Antenna position optimization in movable/fluid antenna systems was studied in \cite{Zhu2023} and \cite{Hu2024}, where the antennas can be shifted on a line segment. The authors in \cite{Gao2024} proposed a joint transmitter and receiver design to maximize the weighted minimum signal-to-interference-plus-noise ratio in a multicast multiple-input single-output (MISO) communication system. The channel estimation problem in movable-antenna MIMO systems was investigated in \cite{Zhang2024}, using tensor decomposition. Nevertheless, efficient port selection algorithms at both ends in fluid-MIMO systems (i.e., traditional MIMO but with fluid antennas) are still missing from the literature.  

In this work, we propose low-complexity algorithms for joint transmit and receive port selection to maximize the channel capacity in fluid-MIMO systems, thereby filling the gap in the literature. It is noted that the authors in \cite{Krikidis2024} studied the maximization of signal-to-noise ratio (SNR) using quantum computation. Also, existing approaches on antenna selection techniques in conventional MIMO systems do not capture this challenge, because they consider either only one-side (transmit/receive) antenna selection \cite{Alkhansari2004, Gorokhov2003, Dua2006}, or a performance metric different from the channel capacity, e.g., energy efficiency with SNR achieved by maximal ratio transmission and combining \cite{Jiang2012}. In massive MIMO systems, the antenna selection problem (at the receiver side) was solved by using branch-and-bound \cite{Gao2018} and greedy \cite{Konar2017} algorithms. The main contributions of this letter are summarized as follows: 
\begin{itemize} 
\item First of all, we formulate the joint transmit and receive fluid antenna port selection problem, and then we provide a joint convex relaxation (JCR) problem by using an upper bound on the channel capacity and taking advantage of the binary optimization variables. As far as we know, the JCR approach is proposed for the first time in both traditional and modern (fluid) MIMO systems.\footnote{In the classical work of receive antenna selection using convex optimization \cite{Dua2006}, the channel capacity is \emph{already} concave and the binary variables are relaxed to be in $[0,1]$. This approach does not suffice for the joint transmit and receive antenna selection, since the channel capacity is \emph{not} concave anymore. For more details on the new convex relaxation technique, see Section~\ref{section:Joint_convex_relaxation}.}

\item Moreover, we design and analyze two optimization algorithms based on the JCR problem, whose complexity is much lower than that of the exhaustive search method. 

\item Finally, numerical results show that the proposed algorithms significantly outperform two baseline schemes, the random port selection and the conventional MIMO setup.   
\end{itemize}

The remainder of this letter is organized as follows. Section~\ref{section:System_model} presents the system model, while Section~\ref{section:Problem_formulation} formulates the discrete optimization problem. Section~\ref{section:Joint_convex_relaxation} provides the joint convex relaxation problem, and Section~\ref{section:Optimization_algorithms} analyzes the optimization algorithms. Furthermore, Section~\ref{section:Numerical_results} presents some numerical results, and Section~\ref{section:Conclusion} concludes the paper. 

\textit{Mathematical notation}:  $\mathbf{I}_N$ is the $N \times N$ identity matrix, and $\mathbf{O}_{N \times M}$ is the $N \times M$ zero matrix. $\mathbf{1}_N$ and $\mathbf{0}_N$ represent the $N$-dimensional all-ones and zero vectors, respectively. Furthermore, $\operatorname{tr}(\cdot)$, $(\cdot)^\top$ and $(\cdot)^H$ stand for the matrix trace, transpose and conjugate-transpose operations, respectively, while $\operatorname{diag}(\mathbf{x})$ is the diagonal matrix having the elements of vector $\mathbf{x}$ on its main diagonal. The determinant is denoted by $\det(\cdot)$, the natural logarithm by $\log(\cdot)$, the ceiling function by $\lceil \cdot \rceil$, and the expectation by $\mathbb{E}(\cdot)$. The symbols $\coloneqq$ and $\eqqcolon$ indicate that the quantity on the side of the colon is defined by the quantity on the other side. The Frobenius norm of a matrix $\mathbf{A} = [a_{i,j}]_{}$ is defined as $\|\mathbf{A}\|_F \coloneqq  \sqrt{\sum_{i,j} |a_{i,j}|^2} = \sqrt{\operatorname{tr}(\mathbf{A} {\mathbf{A}}^H)}$.

\vspace{-4mm}
\section{System Model} \label{section:System_model}

\begin{figure}[!t]
\centering
\includegraphics[width=3.2in]{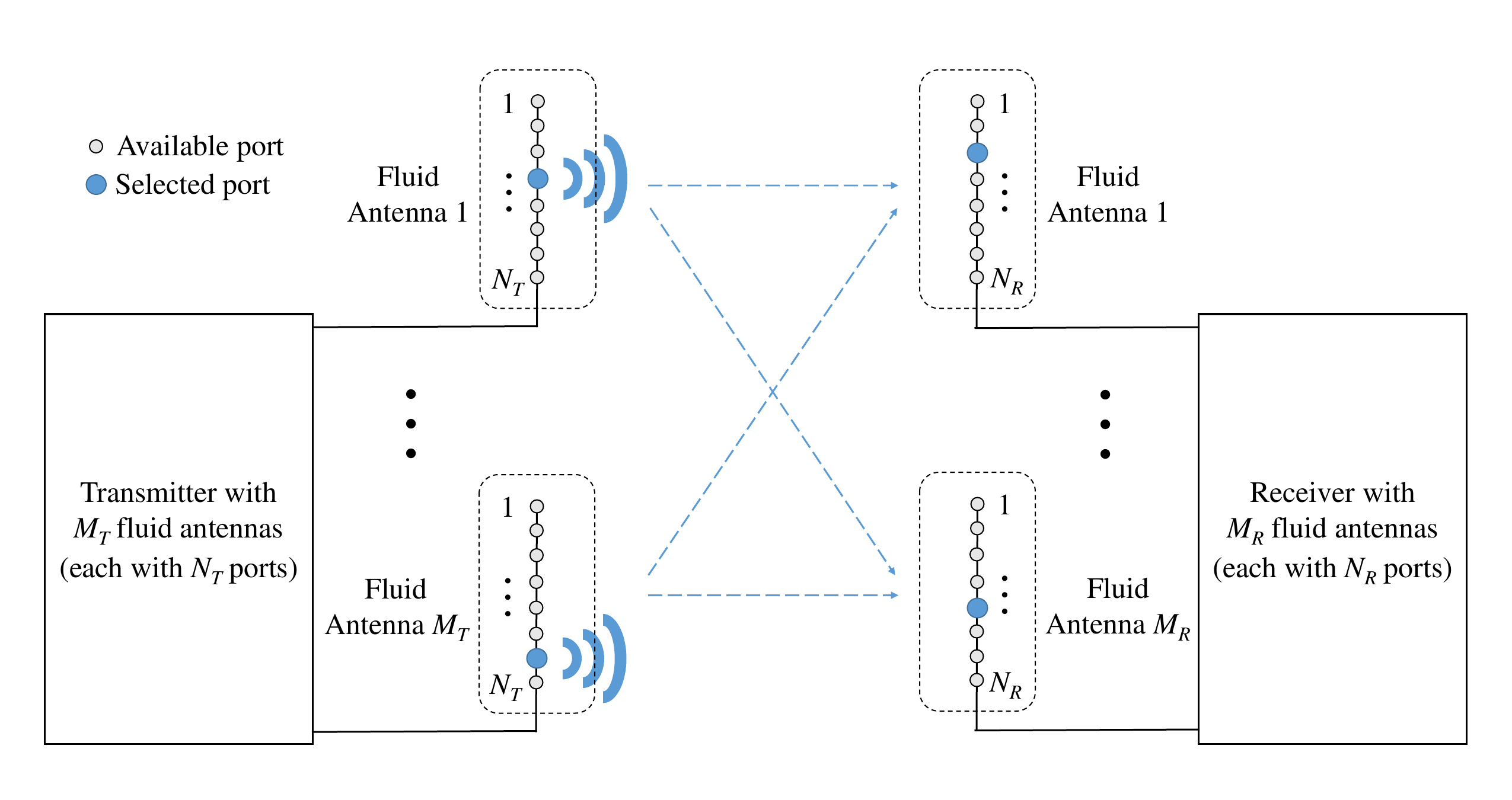} 
\caption{Fluid-MIMO system consisting of a transmitter and a receiver equipped with multiple fluid antennas.}
\label{fig:Fluid-MIMO_system}
\vspace{-5mm}
\end{figure}

We consider a point-to-point fluid-MIMO communication system as shown in Fig.~\ref{fig:Fluid-MIMO_system}, where the transmitter and receiver are equipped with $M_T$ ($\mathcal{M}_T \coloneqq \{1,\dots,M_T\}$) and $M_R$ ($\mathcal{M}_R \coloneqq \{1,\dots,M_R\}$) fluid antennas, respectively. Each fluid antenna at the transmitter and receiver has $N_T$ ($\mathcal{N}_T \coloneqq \{1,\dots,N_T\}$) and $N_R$ ($\mathcal{N}_R \coloneqq \{1,\dots,N_R\}$) available ports/positions, respectively; all ports in a given fluid antenna share a common radio-frequency (RF) chain \cite{Wong2021}. In every time slot, \emph{exactly one} port is selected per transmit/receive fluid antenna; this operation is performed by a central controller. We also assume flat fading and perfect channel state information.\footnote{Note that fluid antennas can use either liquids or reconfigurable pixels \cite{Wong2021}. In liquid-based FAS, the switching time between ports becomes important when the number of ports is very large. In RF pixel-based FAS, however, the switching time is insignificant regardless of the number of ports. In this work, the port switching time is \emph{negligible} compared to the channel coherence time.}

The separation distance between the transmit/receive fluid antennas is sufficiently large (normally, greater than or equal to $\lambda/2$, where $\lambda$ is the wavelength), so that their spatial correlation and mutual coupling are negligible; the channels of distinct fluid antennas are therefore \emph{independent} as in classical MIMO systems. However, \emph{non-negligible} spatial correlation can exist between the ports of a fluid antenna.\footnote{Mutual coupling between the ports of a fluid antenna is not considered in this work, because: a) it does not occur at all in a liquid-based fluid antenna where a single port is selected, and b) it may exist in an RF pixel-based fluid antenna, but it can be mitigated using matching networks \cite[Appendix III]{New2023}.}    

The $(M_R N_R) \times (M_T N_T)$ \emph{overall channel matrix} is denoted by $\mathbf{G} = \left[\mathbf{G}^{(i,j)}\right]_{\substack{i \in \mathcal{M}_R \\ j \in \mathcal{M}_T}}$, where the $N_R \times N_T$ sub-matrix $\mathbf{G}^{(i,j)}$ is given by $\mathbf{G}^{(i,j)} = \left[g^{(i,j)}_{n,k}\right]_{\substack{n \in \mathcal{N}_R \\ k \in \mathcal{N}_T}}$, with $g^{(i,j)}_{n,k}$ being the normalized (i.e., $\mathbb{E}(|g^{(i,j)}_{n,k}|^2) = 1$) channel coefficient between the $n^\text{th}$ port of the $i^\text{th}$ receive fluid antenna and the $k^\text{th}$ port of the $j^\text{th}$ transmit fluid antenna. Let $\widetilde{\mathbf{G}}$ be the $M_R \times M_T$ \emph{effective channel matrix} resulting from the selection of fluid antenna ports. In particular, $\widetilde{\mathbf{G}}$ can be obtained from $\mathbf{G}$ by appropriately selecting subsets of its rows and columns. Now, the received signal $\mathbf{r} = [r_1,\dots,r_{M_R}]^\top$ can be written as $\mathbf{r} = {\delta} \sqrt{P/{M_T}} {\widetilde{\mathbf{G}}} \mathbf{s} + \mathbf{w}$, where $\delta^2$ is the large-scale path loss, $P$ is the total transmit power (divided evenly between the transmit antennas), $\mathbf{s} = [s_1,\dots,s_{M_T}]^\top$ is the transmitted signal ($\mathbb{E}(\mathbf{s} \mathbf{s}^H)=\mathbf{I}_{M_T}$), and $\mathbf{w} = [w_1,\dots,w_{M_R}]^\top$ is the additive white Gaussian noise (AWGN) at the receiver with $\mathbb{E}(\mathbf{w} \mathbf{w}^H)={\sigma_w^2}\mathbf{I}_{M_R}$. Assuming that the entries of $\widetilde{\mathbf{G}}$ are \emph{independent} complex-Gaussian random variables, the channel capacity is given by the well-known formula \cite{Telatar1999} 
\begin{equation} \label{equation:Channel_capacity_original} 
C(\widetilde{\mathbf{G}}) = \log_2 \det(\mathbf{I}_{M_R} + \rho {\widetilde{\mathbf{G}}} {\widetilde{\mathbf{G}}^H}) = \log_2 \det(\mathbf{I}_{M_T} + \rho {\widetilde{\mathbf{G}}^H} {\widetilde{\mathbf{G}}}) , 
\end{equation}
where $\rho = \overline{\gamma}/{M_T}$, with $\overline{\gamma} = {{\delta^2} P}/{\sigma_w^2}$ being the average SNR at each receive fluid antenna.

\vspace{-2mm}
\section{Problem Formulation} \label{section:Problem_formulation}

First of all, we introduce the binary decision variables $\mathbf{x} = [\mathbf{x}^{(i)}]_{i \in \mathcal{M}_R} \in \{0,1\}^{M_R N_R}$ and $\mathbf{y} = [\mathbf{y}^{(j)}]_{j \in \mathcal{M}_T} \in \{0,1\}^{M_T N_T}$, where $\mathbf{x}^{(i)} = [x^{(i)}_1,\dots,x^{(i)}_{N_R}]$ and $\mathbf{y}^{(j)} = [y^{(j)}_1,\dots,y^{(j)}_{N_T}]$. In particular, each variable $x^{(i)}_n$ or $y^{(j)}_k$ is equal to $1$ if the corresponding port is selected, and $0$ otherwise. Now, let us define the $(M_R N_R) \times (M_T N_T)$ matrix 
\begin{equation}
\mathbf{Q} \coloneqq \mathbf{X} \mathbf{G} \mathbf{Y} = \left[ \mathbf{Q}^{(i,j)} \right]_{\substack{i \in \mathcal{M}_R \\ j \in \mathcal{M}_T}}  ,
\end{equation} 
where $\mathbf{X} = \operatorname{diag}(\mathbf{x})$, $\mathbf{Y} = \operatorname{diag}(\mathbf{y})$, $\mathbf{Q}^{(i,j)} = \mathbf{X}^{(i)} \mathbf{G}^{(i,j)} \mathbf{Y}^{(j)} = \left[{x^{(i)}_n} {g^{(i,j)}_{n,k}} {y^{(j)}_k}\right]_{\substack{n \in \mathcal{N}_R \\ k \in \mathcal{N}_T}}$, $\mathbf{X}^{(i)} = \operatorname{diag}(\mathbf{x}^{(i)})$, and $\mathbf{Y}^{(j)} = \operatorname{diag}(\mathbf{y}^{(j)})$. In essence, the diagonal matrices $\mathbf{X}$ and $\mathbf{Y}$ result in row and column selection of $\mathbf{G}$, respectively.

\begin{proposition}
With a slight abuse of notation, we can rewrite the channel capacity given by \eqref{equation:Channel_capacity_original} as follows
\begin{equation} \label{equation:Channel_capacity}
C(\mathbf{x},\mathbf{y}) \coloneqq C(\widetilde{\mathbf{G}}) = \log_2 \det (\mathbf{I}_{M_R N_R} + \rho \mathbf{Q} {\mathbf{Q}^H})  . 
\end{equation}
\end{proposition}

\begin{IEEEproof}
First of all, matrix $\mathbf{Q}$ can be also expressed as $\mathbf{Q} = \mathbf{P}_\text{r} {\widehat{\mathbf{G}}} \mathbf{P}_\text{c}$, where $\mathbf{P}_\text{r}$ and $\mathbf{P}_\text{c}$ are appropriately chosen permutation matrices of dimension $(M_R N_R) \times (M_R N_R)$ and $(M_T N_T) \times (M_T N_T)$, respectively, and 
\begin{equation}
\widehat{\mathbf{G}} = \begin{bmatrix} \widetilde{\mathbf{G}} & \mathbf{O}_{M_R \times d_2} \\ \mathbf{O}_{d_1 \times M_T} & \mathbf{O}_{d_1 \times d_2} \end{bmatrix}
\end{equation}
with $d_1 = M_R (N_R - 1)$ and $d_2 = M_T (N_T - 1)$. 

Since for any permutation matrix $\mathbf{P}$ it holds that $\mathbf{P}^H = \mathbf{P}^\top = \mathbf{P}^{-1}$, we obtain $\mathbf{Q} {\mathbf{Q}^H} = {\mathbf{P}_\text{r}} {\widehat{\mathbf{G}}} {\widehat{\mathbf{G}}^H} {\mathbf{P}_\text{r}^H}$. In addition, for any matrices $\mathbf{A}$ ($N \times M$) and $\mathbf{B}$ ($M \times N$), we have the matrix identity: $\det(\mathbf{I}_N + \mathbf{A}\mathbf{B}) = \det(\mathbf{I}_M + \mathbf{B}\mathbf{A})$. As a result, 
\begin{equation}
\det(\mathbf{I}_{M_R N_R} + \rho \mathbf{Q} {\mathbf{Q}^H})= \det(\mathbf{I}_{M_R N_R} + \rho {\widehat{\mathbf{G}}} {\widehat{\mathbf{G}}^H}) .
\end{equation}

Moreover, ${\widehat{\mathbf{G}}^H} = \begin{bmatrix} \widetilde{\mathbf{G}}^H & \mathbf{O}_{M_T \times d_1} \\ \mathbf{O}_{d_2 \times M_R}  & \mathbf{O}_{d_2 \times d_1} \end{bmatrix}$ and ${\widehat{\mathbf{G}}} {\widehat{\mathbf{G}}^H} = \begin{bmatrix} \widetilde{\mathbf{G}} {\widetilde{\mathbf{G}}^H} & \mathbf{O}_{M_R \times d_1} \\ \mathbf{O}_{d_1 \times M_R} & \mathbf{O}_{d_1 \times d_1} \end{bmatrix}$, thus 
\begin{equation}   
\mathbf{I}_{M_R N_R} + \rho {\widehat{\mathbf{G}}} {\widehat{\mathbf{G}}^H} = \begin{bmatrix}
  \mathbf{I}_{M_R} + \rho \widetilde{\mathbf{G}} {\widetilde{\mathbf{G}}^H} & \mathbf{O}_{M_R \times d_1} \\
  \mathbf{O}_{d_1 \times M_R}  & \mathbf{I}_{d_1} 
\end{bmatrix}            .
\end{equation}
For any matrices $\mathbf{A}$ ($N \times N$), $\mathbf{B}$ ($N \times M$), and $\mathbf{D}$ ($M \times M$) we have the following identity of block matrix determinant: $\det\left( \begin{bmatrix} \mathbf{A} & \mathbf{B} \\ \mathbf{O}_{M \times N} & \mathbf{D} \end{bmatrix} \right) = \det(\mathbf{A}) \det(\mathbf{D})$.
Therefore, based on the previous equations we obtain  
\begin{equation}
\log_2 \det(\mathbf{I}_{M_R N_R} + \rho \mathbf{Q} {\mathbf{Q}^H}) =  C(\widetilde{\mathbf{G}}) ,
\end{equation}
because $\det(\mathbf{I}_{d_1}) = 1$. 
\end{IEEEproof}

Now, we can formulate a \emph{binary optimization problem} to select the ports that maximize the channel capacity, i.e.,     
\begin{subequations} \label{equation:Binary_problem}
\begin{alignat}{3}
 C^* \coloneqq \, & \mathop {\max}\limits_{\mathbf{x},\mathbf{y}} & \quad & C(\mathbf{x},\mathbf{y})   \\
  & ~\text{s.t.} & & \sum_{n \in \mathcal{N}_R} x^{(i)}_n = 1, \;\; \forall i \in \mathcal{M}_R ,  \label{equation:Sum_x_constraint} \\ 
  & & &  \sum_{k \in \mathcal{N}_T} y^{(j)}_k = 1, \;\; \forall j \in \mathcal{M}_T , \label{equation:Sum_y_constraint} \\ 
  & & &  \mathbf{x} \in \{0,1\}^{M_R N_R} , \;\;  \mathbf{y} \in \{0,1\}^{M_T N_T}  .  
\end{alignat}
\end{subequations}
The first two equality-constraints ensure that exactly one port is selected for each fluid antenna. Note that problem \eqref{equation:Binary_problem} seems very difficult to solve in polynomial time.

\section{Joint Convex Relaxation} \label{section:Joint_convex_relaxation}

In this section, we use the following result in order to upper bound the channel capacity, which is \emph{not} jointly concave.  

\begin{proposition} \label{proposition:log_det_inequality}
If $\mathbf{B}$ is a positive definite $N \times N$ matrix, then we have 
\begin{equation}
\log \det(\mathbf{B}) \leq \operatorname{tr}(\mathbf{B}-\mathbf{I}_N) .
\end{equation} 
\end{proposition}

\begin{IEEEproof}
Let $\{\lambda_n\}_{n \in \mathcal{N}}$, where $\mathcal{N} \coloneqq \{1,\dots,N\}$, be the eigenvalues of matrix $\mathbf{B}$. Since $\mathbf{B}$ is positive definite, $\lambda_n > 0$ for all $n \in \mathcal{N}$. Now, we can write $\log \det(\mathbf{B}) = \log (\prod_{n \in \mathcal{N}} {\lambda_n}) = \sum_{n \in \mathcal{N}} {\log \lambda_n}$. In addition, it holds that $\log x \leq x - 1$ for all $x>0$. Therefore, $\log \det(\mathbf{B}) \leq \sum_{n \in \mathcal{N}} (\lambda_n - 1) = \operatorname{tr}(\mathbf{B}-\mathbf{I}_N)$.
\end{IEEEproof}

Proposition \ref{proposition:log_det_inequality} with $\mathbf{B} = \mathbf{I}_{M_R N_R} + \rho \mathbf{Q} {\mathbf{Q}^H}$ implies that\footnote{The matrix $\mathbf{R} \coloneqq \mathbf{Q} {\mathbf{Q}^H}$ is positive semi-definite, so its eigenvalues $\lambda_n^{(\mathbf{R})} \geq 0$ for all $n$. The eigenvalues of matrix $\mathbf{B}$ are $\lambda_n^{(\mathbf{B})} = 1 + \rho \lambda_n^{(\mathbf{R})} \geq 1 > 0$ for all $n$, thus $\mathbf{B}$ is positive definite; note that $\rho \geq 0$.}  
\begin{equation} \label{equation:Channel_capacity_upper_bound} 
C(\mathbf{x},\mathbf{y}) \leq \tfrac{\rho}{\log 2} \operatorname{tr}(\mathbf{Q} {\mathbf{Q}^H}) = \tfrac{\rho}{\log 2} \|\mathbf{Q}\|_F^2 .
\end{equation}
Now, we make a \emph{key observation}: Due to the fact that $\mathbf{x}$ and $\mathbf{y}$ are binary vectors, we have $(x^{(i)}_n)^2 = x^{(i)}_n$, $(y^{(j)}_k)^2 = y^{(j)}_k$, and ${x^{(i)}_n} {y^{(j)}_k} = \min(x^{(i)}_n,y^{(j)}_k)$. As a result, 
\begin{equation} \label{equation:U_definition} 
\begin{split} 
\|\mathbf{Q}\|_F^2 & \coloneqq \sum_{i \in \mathcal{M}_R} \sum_{n \in \mathcal{N}_R} \sum_{j \in \mathcal{M}_T} \sum_{k \in \mathcal{N}_T} {|{x^{(i)}_n} {g^{(i,j)}_{n,k}} {y^{(j)}_k}|^2}  \\ 
& = \sum_{i \in \mathcal{M}_R} \sum_{n \in \mathcal{N}_R} \sum_{j \in \mathcal{M}_T} \sum_{k \in \mathcal{N}_T} {|{g^{(i,j)}_{n,k}}|^2 \min(x^{(i)}_n,y^{(j)}_k)}   \\
& \eqqcolon U(\mathbf{x},\mathbf{y}) , \;\;  \forall (\mathbf{x},\mathbf{y}) \in \{0,1\}^{M_R N_R} \times \{0,1\}^{M_T N_T} . 
\end{split} 
\end{equation} 
Note that $U(\mathbf{x},\mathbf{y})$ is \emph{jointly concave in $\mathbf{x}$ and $\mathbf{y}$}, because $\min(x^{(i)}_n,y^{(j)}_k)$ is a concave function \cite{Boyd2004}.  

Subsequently, we formulate a \emph{joint convex relaxation (JCR) problem} by relaxing the objective and the binary constraints: 
\begin{subequations} \label{equation:JCR_problem}
\begin{alignat}{3}
 U^* \coloneqq \,  & \mathop {\max}\limits_{\mathbf{x},\mathbf{y}} & \quad & U(\mathbf{x},\mathbf{y})   \\
  & ~\text{s.t.} & & \eqref{equation:Sum_x_constraint} , \;\; \eqref{equation:Sum_y_constraint} ,  \\  
  & & &  \mathbf{x} \in [0,1]^{M_R N_R} , \;\;  \mathbf{y} \in [0,1]^{M_T N_T}  .  
\end{alignat}
\end{subequations}
Observe that $C^* \leq \tfrac{\rho}{\log 2} U^*$, due to \eqref{equation:Channel_capacity_upper_bound} and \eqref{equation:U_definition}, and because $\mathcal{S}_C \subseteq \mathcal{S}_U$, where $\mathcal{S}_C$ and $\mathcal{S}_U$ are the feasible sets of problems \eqref{equation:Binary_problem} and \eqref{equation:JCR_problem}, respectively.

Finally, given that the JCR problem has $\nu = M_R N_R + M_T N_T$ decision variables and $\kappa = M_R + M_T + 2(M_R N_R + M_T N_T) = \Theta(M_R N_R + M_T N_T)$ constraints, a globally optimal solution can be computed in $T_{\text{JCR}} = O((M_R N_R + M_T N_T)^{3.5})$ time using an interior-point method, which is expected to be $O((M_R N_R + M_T N_T)^{3})$ in practice \cite{Boyd2004}.

\section{Optimization Algorithms} \label{section:Optimization_algorithms}

Given two matrices $\mathbf{A}$ ($N \times M$) and $\mathbf{B}$ ($M \times N$), the computation of $\mathbf{A} \mathbf{B}$ (matrix multiplication) requires $O(N^2 M)$ arithmetic operations. In addition, the determinant of an $N \times N$ matrix can be computed in $O(N^3)$ time using, for example, lower-upper (LU) decomposition. Therefore, the computation of channel capacity given by \eqref{equation:Channel_capacity_original} requires either $O({M_R^2} {M_T} + {M_R^3})$ or $O({M_T^2} {M_R} + {M_T^3})$ time depending on the formula used. By selecting the fastest between the two, we have $T_C = O((\min(M_R , M_T))^2 (M_R + M_T))$ complexity.

\subsection{Exhaustive Search} 

The exhaustive search method generates all feasible solutions and selects that with the highest channel capacity, thereby achieving the \emph{global optimum}. Since the number of feasible solutions is $(N_R)^{M_R} (N_T)^{M_T}$, its running time is $O((N_R)^{M_R} (N_T)^{M_T} T_C)$, thus having \emph{exponential complexity}.

\subsection{Joint Convex Relaxation \& Reduced Exhaustive Search}

\begin{algorithm}[!t]
\caption{Joint Convex Relaxation \& Reduced Exhaustive Search (JCR\&RES)} \label{algorithm:JCR_RES}
\footnotesize 
\begin{algorithmic}[1]
\State Solve the joint convex relaxation problem \eqref{equation:JCR_problem} to find an optimal solution $({\widehat{\mathbf{x}}},{\widehat{\mathbf{y}}}) \in [0,1]^{M_R N_R} \times [0,1]^{M_T N_T}$.  
\State $\widetilde{N}_R \leftarrow \lceil \log_2 (N_R+1) \rceil$, $\widetilde{N}_T \leftarrow \lceil \log_2 (N_T+1) \rceil$ 
\State For each receive fluid antenna $i \in \mathcal{M}_R$, choose the $\widetilde{N}_R$ ports with the largest $\widehat{x}^{(i)}_n \in [0,1]$ (among all the $N_R$ ports) and let $\widetilde{\mathcal{N}}_R^{(i)}$ be the set of these ports.
\State For each transmit fluid antenna $j \in \mathcal{M}_T$, choose the $\widetilde{N}_T$ ports with the largest $\widehat{y}^{(j)}_k \in [0,1]$ (among all the $N_T$ ports) and let $\widetilde{\mathcal{N}}_T^{(j)}$ be the set of these ports. 
\State Let $\mathbf{G}_{\text{new}}$ be the $(M_R \widetilde{N}_R)\times(M_T \widetilde{N}_T)$ sub-matrix of $\mathbf{G}$ obtained by selecting the subset $\bigcup_{i \in \mathcal{M}_R} \widetilde{\mathcal{N}}_R^{(i)}$ of its rows and the subset $\bigcup_{j \in \mathcal{M}_T} \widetilde{\mathcal{N}}_T^{(j)}$ of its columns. 
\State Apply the exhaustive search method to solve the \emph{reduced-dimension} problem \eqref{equation:Binary_problem} by replacing $\mathbf{G} \mapsto \mathbf{G}_{\text{new}}$, $N_R \mapsto \widetilde{N}_R$, and $N_T \mapsto \widetilde{N}_T$. Let $(\mathbf{x}^{\star},\mathbf{y}^{\star}) \in \{0,1\}^{M_R \widetilde{N}_R} \times \{0,1\}^{M_T \widetilde{N}_T}$ be one of its optimal solutions.  
\State $\widetilde{\mathbf{x}} \leftarrow \mathbf{0}_{M_R N_R}$, $\widetilde{\mathbf{x}} (\bigcup_{i \in \mathcal{M}_R} \widetilde{\mathcal{N}}_R^{(i)}) \leftarrow \mathbf{x}^{\star}$
\State $\widetilde{\mathbf{y}} \leftarrow \mathbf{0}_{M_T N_T}$, $\widetilde{\mathbf{y}} (\bigcup_{j \in \mathcal{M}_T} \widetilde{\mathcal{N}}_T^{(j)}) \leftarrow \mathbf{y}^{\star}$
\State \textbf{return} $(\widetilde{\mathbf{x}},\widetilde{\mathbf{y}})$
\end{algorithmic}
\end{algorithm} 

The JCR\&RES method is presented in Algorithm~\ref{algorithm:JCR_RES}. After solving the JCR problem, the algorithm forms a subset of the available ports having the largest fractional solutions, and then applies the exhaustive search to solve a problem of \emph{much smaller dimension} than the original one. Since $\lceil \log_2 (N+1) \rceil = \Theta(\log_2 N)$, Algorithm~\ref{algorithm:JCR_RES} requires $O(T_{\text{JCR}} + (\log_2 N_R)^{M_R} (\log_2 N_T)^{M_T} T_C)$ time. Its complexity is therefore exponential, but much lower than that of exhaustive search. As a consequence, JCR\&RES is suitable for practical applications with a small to moderate number of ports.

\subsection{Joint Convex Relaxation \& Alternating Optimization}

\begin{algorithm}[!t]
\caption{Joint Convex Relaxation \& Alternating Optimization (JCR\&AO)} \label{algorithm:JCR_AO}
\footnotesize 
\begin{algorithmic}[1]
\State Select a convergence tolerance $\epsilon >0$ and a maximum number of iterations $I \geq 1$ for alternating optimization. 
\State Solve the joint convex relaxation problem \eqref{equation:JCR_problem} to find an optimal solution $({\widehat{\mathbf{x}}},{\widehat{\mathbf{y}}}) \in [0,1]^{M_R N_R} \times [0,1]^{M_T N_T}$.  
\State $\widetilde{\mathbf{x}} \leftarrow \mathbf{0}_{M_R N_R}$, $\widetilde{\mathbf{y}} \leftarrow \mathbf{0}_{M_T N_T}$

\ForAll{$i \in \mathcal{M}_R$}  $n^{\star} \leftarrow {\arg \max}_{n \in \mathcal{N}_R} \{\widehat{x}^{(i)}_n\}$, $\widetilde{x}^{(i)}_{n^{\star}} \leftarrow 1$  
\EndFor

\ForAll{$j \in \mathcal{M}_T$}  $k^{\star} \leftarrow {\arg \max}_{k \in \mathcal{N}_T} \{\widehat{y}^{(j)}_k\}$, $\widetilde{y}^{(j)}_{k^{\star}} \leftarrow 1$   
\EndFor

\State $C_{\text{old}} \leftarrow 0$, $C_{\text{new}} \leftarrow C(\widetilde{\mathbf{x}},\widetilde{\mathbf{y}})$, $C_{\text{best}} \leftarrow C_{\text{new}}$, $\ell \leftarrow 0$   
\While{$(|C_{\text{new}} - C_{\text{old}}| > |C_{\text{old}}|\epsilon)$ \& $(\ell < I)$}
	\State $C_{\text{old}} \leftarrow C_{\text{new}}$
	
	\ForAll{$i \in \mathcal{M}_R$} 
		\State $\widetilde{x}^{(i)}_{n'} \leftarrow 0$ for all $n' \in \mathcal{N}_R$ 
		\ForAll{$n \in \mathcal{N}_R$} 
			\If{$n \geq 2$}  $\widetilde{x}^{(i)}_{n-1} \leftarrow 0$, $\widetilde{x}^{(i)}_n \leftarrow 1$ \textbf{else} $\widetilde{x}^{(i)}_1 \leftarrow 1$    
			\EndIf 
			
			\If{$C(\widetilde{\mathbf{x}},\widetilde{\mathbf{y}}) \geq C_{\text{best}}$} $C_{\text{best}} \leftarrow C(\widetilde{\mathbf{x}},\widetilde{\mathbf{y}})$, $n_{\text{best}} \leftarrow n$ 
			\EndIf	
		\EndFor 
		\State $\widetilde{x}^{(i)}_{N_R} \leftarrow 0$, $\widetilde{x}^{(i)}_{n_{\text{best}}} \leftarrow 1$ 
	\EndFor
	
	\ForAll{$j \in \mathcal{M}_T$} 
		\State $\widetilde{y}^{(j)}_{k'} \leftarrow 0$ for all $k' \in \mathcal{N}_T$ 
		\ForAll{$k \in \mathcal{N}_T$} 
			\If{$k \geq 2$}  $\widetilde{y}^{(j)}_{k-1} \leftarrow 0$, $\widetilde{y}^{(j)}_k \leftarrow 1$ \textbf{else}  $\widetilde{y}^{(j)}_1 \leftarrow 1$     
			\EndIf  
			
			\If{$C(\widetilde{\mathbf{x}},\widetilde{\mathbf{y}}) \geq C_{\text{best}}$}  $C_{\text{best}} \leftarrow C(\widetilde{\mathbf{x}},\widetilde{\mathbf{y}})$, $k_{\text{best}} \leftarrow k$  
			\EndIf	
		\EndFor 
		\State $\widetilde{y}^{(j)}_{N_T} \leftarrow 0$, $\widetilde{y}^{(j)}_{k_{\text{best}}} \leftarrow 1$ 
	\EndFor
	 
	\State $C_{\text{new}} \leftarrow C_{\text{best}}$, $\ell \leftarrow \ell + 1$ 
\EndWhile

\State \textbf{return} $(\widetilde{\mathbf{x}},\widetilde{\mathbf{y}})$
\end{algorithmic}
\end{algorithm}

The JCR\&AO method is given in Algorithm~\ref{algorithm:JCR_AO}. Based on the (fractional) solution of the JCR problem, the algorithm initially selects the port with the largest fractional solution for each fluid antenna (steps 3-11), and then improves the binary solution using alternating optimization (AO) in steps 12-44. In every AO-iteration, we successively switch to the best (highest-capacity) port for each fluid antenna at the receiver (steps 15-28) and transmitter (steps 29-42), given that all selected ports in the remaining fluid antennas are fixed. The AO technique produces a non-decreasing sequence of channel capacities, i.e., $C(\widetilde{\mathbf{x}}^{(\ell + 1)},\widetilde{\mathbf{y}}^{(\ell + 1)}) \geq C(\widetilde{\mathbf{x}}^{(\ell)},\widetilde{\mathbf{y}}^{(\ell)})$, for all $\ell \geq 0$, and because $C(\mathbf{x},\mathbf{y})$ is upper bounded, the convergence of JCR\&AO is guaranteed. Moreover, it has \emph{polynomial complexity} in the input size, i.e., $O(T_{\text{JCR}} + I (M_R N_R + M_T N_T) T_C)$, assuming a fixed maximum number of AO-iterations $I$. Therefore, Algorithm~\ref{algorithm:JCR_AO} can be successfully applied in fluid-MIMO systems with a small to large number of ports. Finally, the complexity of all optimization algorithms is summarized in Table \ref{table:Complexity_Performance}.

\begin{table}[!t]
\caption{Computational Complexity of Optimization Algorithms} 
\centering 
\renewcommand{\arraystretch}{1.8}
\begin{threeparttable}
\begin{tabular}{|c|c|}
\hline 
\textbf{Algorithm}\textsuperscript{*} & \textbf{Time Complexity}\textsuperscript{**} \\ \hline 
Exhaustive Search & $O((N_R)^{M_R} (N_T)^{M_T} T_C)$  \\ \hline
\makecell{JCR\&RES \\ (Algorithm \ref{algorithm:JCR_RES})} & $O(T_{\text{JCR}} + (\log_2 N_R)^{M_R} (\log_2 N_T)^{M_T} T_C)$  \\ \hline 
\makecell{JCR\&AO \\ (Algorithm \ref{algorithm:JCR_AO})} & $O(T_{\text{JCR}} + I (M_R N_R + M_T N_T) T_C)$  \\ \hline
\end{tabular} 
\begin{tablenotes}
\item[*] Exhaustive search achieves an optimal solution, whereas JCR\&RES and JCR\&AO return a sub-optimal (feasible) solution. 
\item[**] $T_{\text{JCR}} = O((M_R N_R + M_T N_T)^{3.5})$ is the time complexity of solving the joint convex relaxation problem \eqref{equation:JCR_problem} using interior-point methods. $T_C = O((\min(M_R,M_T))^2 (M_R + M_T))$ is the time required for computing the channel capacity given by \eqref{equation:Channel_capacity_original}.   
\end{tablenotes}
\end{threeparttable}
\label{table:Complexity_Performance}
\vspace{-2mm}
\end{table}

\vspace{-2mm}
\section{Numerical Results} \label{section:Numerical_results}

We generated $100$ independent channel matrices $\mathbf{G}$ and computed the average channel capacity achieved by each algorithm. The available ports in each transmit/receive fluid antenna were supposed to be evenly distributed on a line segment of length $W \lambda$, where $\lambda$ is the wavelength. In addition, the (normalized) channel coefficients of fluid antenna ports were modeled as $g^{(i,j)}_{n,k} = \left( \sqrt{1-(\mu^{(i,j)}_{n,k})^2} {u^{(i,j)}_{n,k}} + {\mu^{(i,j)}_{n,k}} {u^{(i,j)}_0}  \right) + \mathrm{i} \left( \sqrt{1-(\mu^{(i,j)}_{n,k})^2} {v^{(i,j)}_{n,k}} + {\mu^{(i,j)}_{n,k}} {v^{(i,j)}_0} \right)$, where $u^{(i,j)}_{n,k}$, $u^{(i,j)}_0$, $v^{(i,j)}_{n,k}$, ${v^{(i,j)}_0}$ are all independent Gaussian random variables with zero mean and variance $1/2$, ${\mu^{(i,j)}_{n,k}} \in [-1,1]$ is the spatial-correlation parameter given by ${\mu^{(i,j)}_{n,k}} = {\mu_{n,k}} = \tfrac{1}{2} \left[ J_0 \left( \frac{2\pi(n-1)}{N_R-1} W \right) + J_0 \left( \frac{2\pi(k-1)}{N_T-1} W \right) \right]$, with $J_0(\cdot)$ being the zero-order Bessel function of the first kind, and $\mathrm{i} = \sqrt{- 1}$ is the imaginary unit \cite{Wong2021}. Observe that $\mu^{(i,j)}_{n,k} = 1$ when $W=0$, whereas $\mu^{(i,j)}_{n,k} \to 0$ as $W \to \infty$. For simplicity, $M_R = M_T = M$ and $N_R = N_T = N$. Unless otherwise specified, the remaining system parameters were $W=0.5$, $N=10$, and $\overline{\gamma}=5\ \text{dB}$. In JCR\&AO, $\epsilon = 10^{-3}$ and $I = 20$.  



\begin{figure}[!t]
\centering
\includegraphics[width=3.4in]{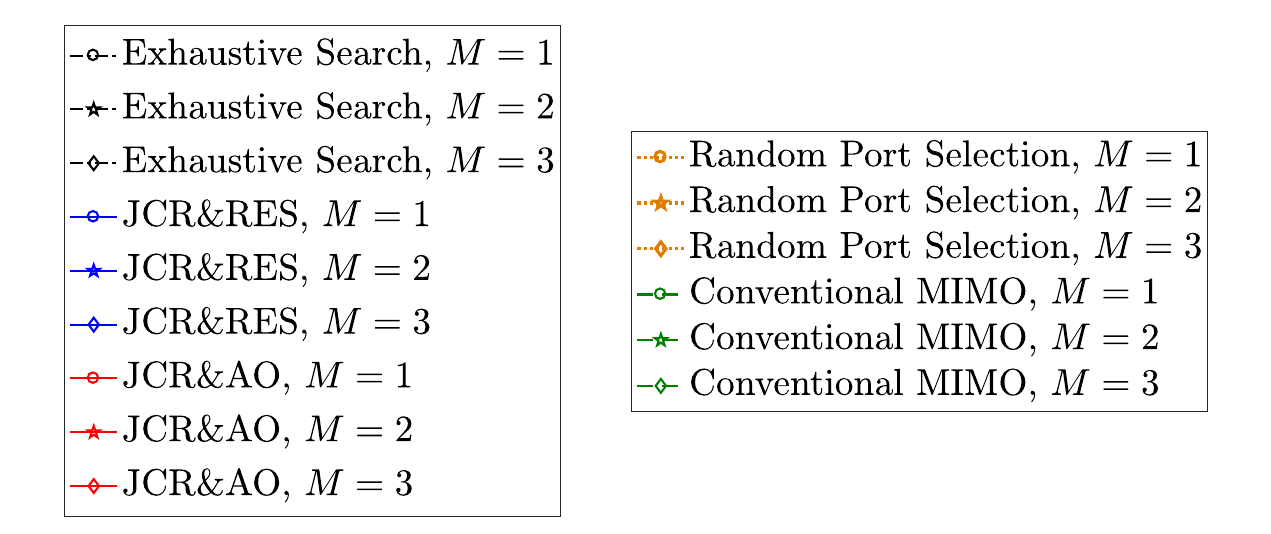} 
\caption{Common legend for the next figures.}
\label{fig:Legend}
\vspace{-2mm}
\end{figure}

\begin{figure}[!t]
\centering
\includegraphics[width=3.2in]{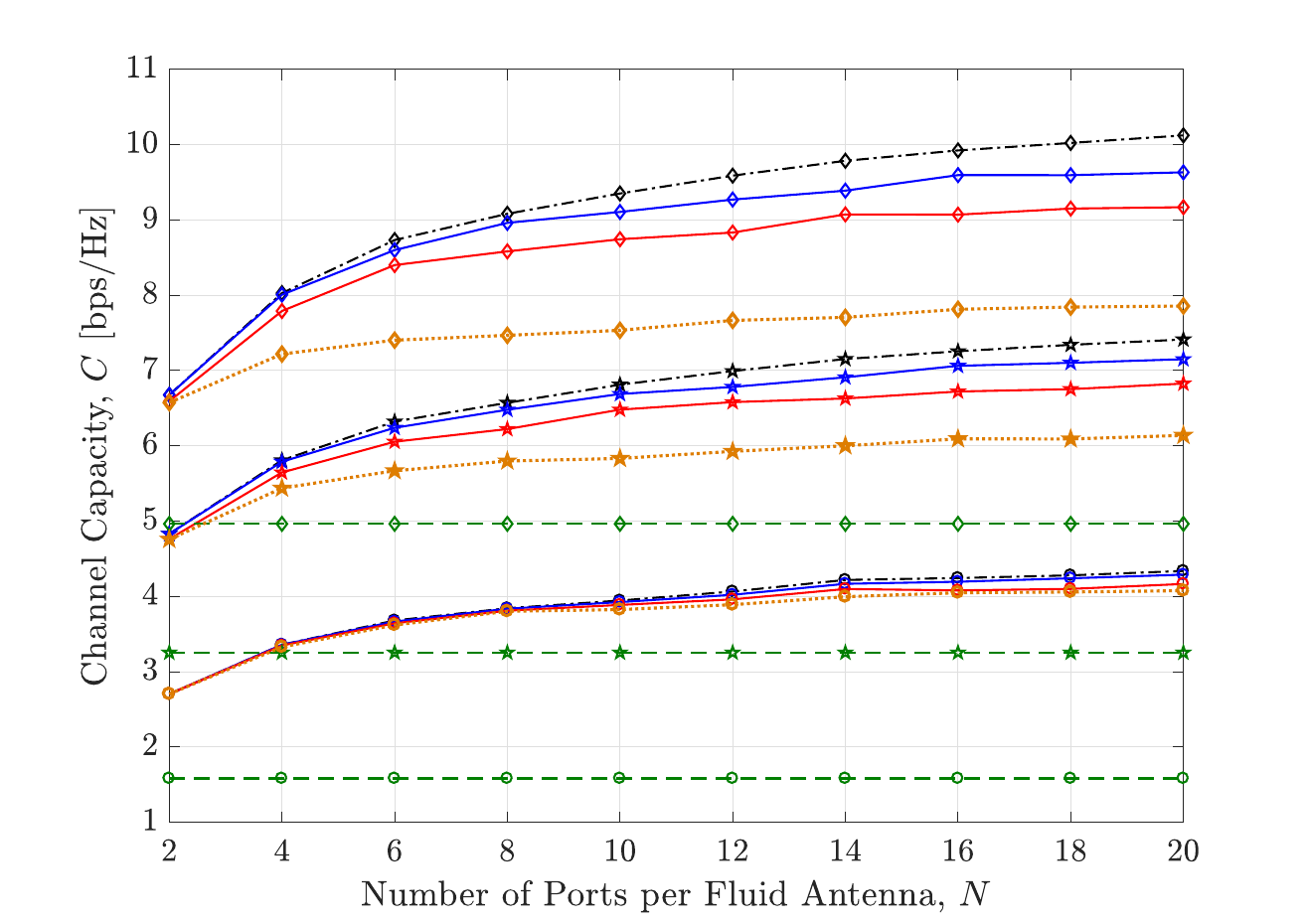} 
\caption{Channel capacity vs the number of ports per fluid antenna. JCR\&AO requires about 1-3 AO-iterations to converge, on average, for all MIMO setups.} 
\label{fig:C_vs_N}
\vspace{-4mm}
\end{figure}

\begin{figure}[!t]
\centering
\includegraphics[width=3.2in]{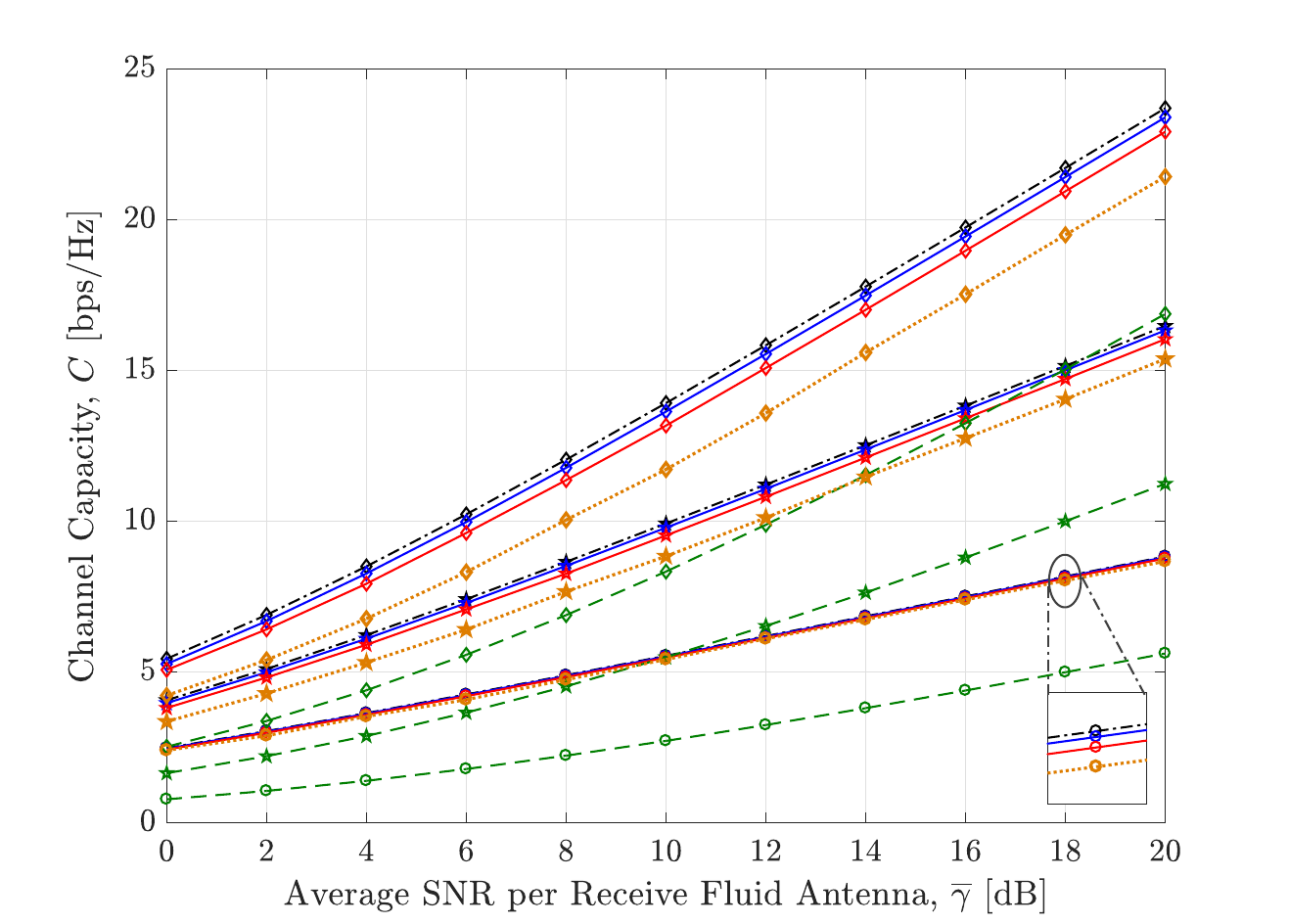} 
\caption{Channel capacity vs the average SNR per receive fluid antenna. JCR\&AO requires around 1-3 AO-iterations to converge, on average, for all MIMO configurations.}
\label{fig:C_vs_SNR}
\vspace{-5mm}
\end{figure}

\begin{figure}[!t]
\centering
\includegraphics[width=3.2in]{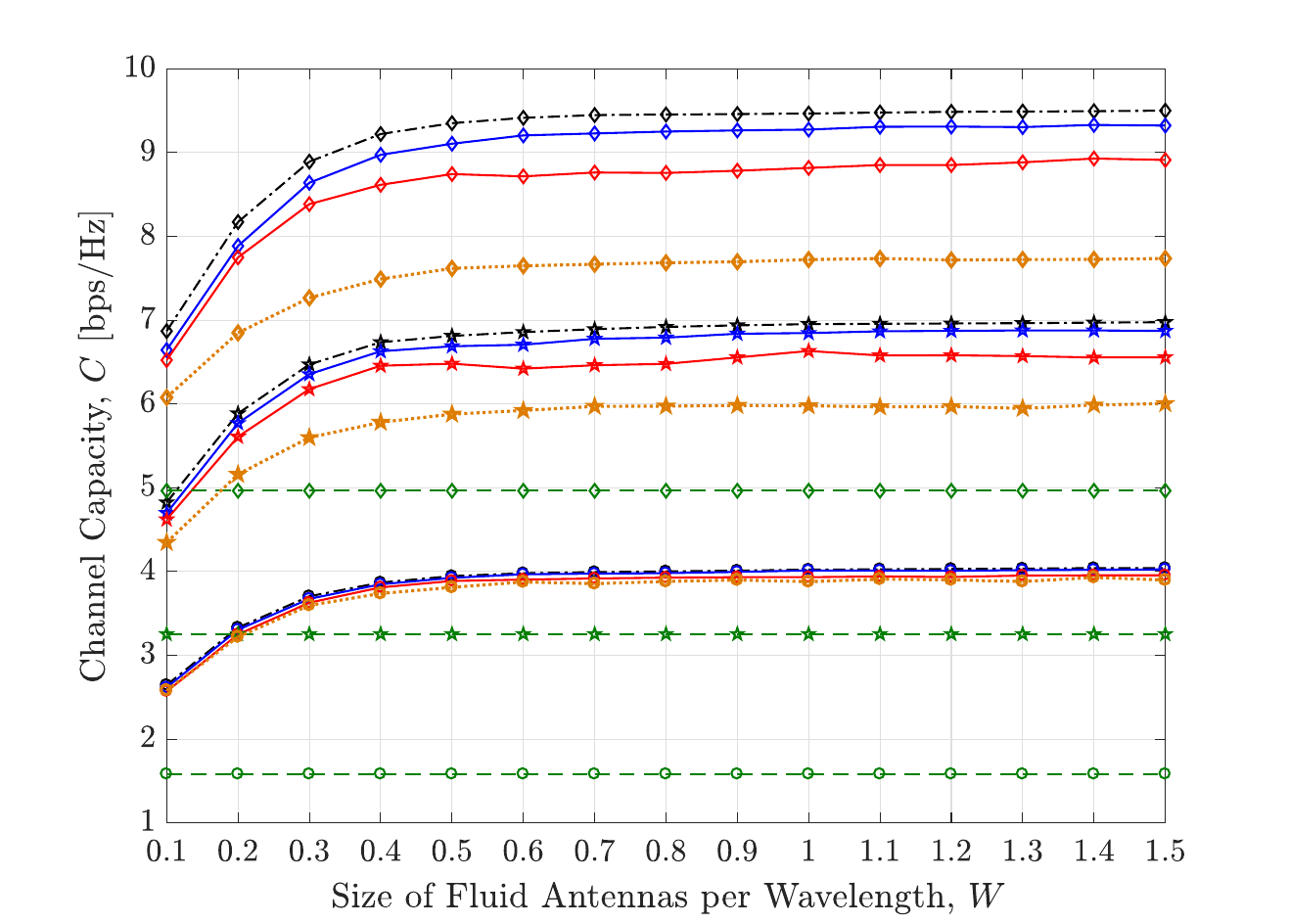} 
\caption{Channel capacity vs the size of fluid antennas per wavelength. JCR\&AO requires approximately 1-3 AO-iterations until convergence, on average, for all MIMO setups.}
\label{fig:C_vs_W}
\vspace{-5mm}
\end{figure}

In addition, two baseline schemes were considered for comparison: a) \textit{Random Port Selection}: we randomly generate $10 N M$ feasible solutions and select that with the highest channel capacity, and b) \textit{Conventional MIMO}: we always select the first port of each fluid antenna. Fig.~\ref{fig:Legend} presents the common legend for the following figures. Note also that JCR\&AO requires about 1-3 AO-iterations to converge, on average, for all MIMO setups (see the captions of Figs.~\ref{fig:C_vs_N}-\ref{fig:C_vs_W}).

First of all, we examine the impact of the number of fluid antenna ports on the channel capacity. In particular, Fig.~\ref{fig:C_vs_N} shows that the optimal channel capacity (achieved by the exhaustive search) increases with the number of ports. Moreover, the proposed algorithms achieve higher performance than the two benchmarks in all simulation scenarios, with Conventional MIMO having the lowest performance. Between JCR\&RES and JCR\&AO, the former exhibits better performance compared to the latter, but in exchange for higher computational complexity (performance-complexity tradeoff). For $N=20$, the approximation ratio (defined as the ratio of the achieved to the optimal channel capacity) per algorithm is: Conventional MIMO $\{36,44,49\}\%$, Random Port Selection $\{94,83,78\}\%$, JCR\&AO $\{96,92,91\}\%$, and JCR\&RES $\{99,96,95\}\%$ for $M =1,2,3$, respectively.

Furthermore, Fig.~\ref{fig:C_vs_SNR} presents the channel capacity against the average SNR at each receive fluid antenna. We can make similar observations about the performance of algorithms as in Fig.~\ref{fig:C_vs_N}. Interestingly, JCR\&RES and JCR\&AO remain very close to the global optimum for low to high SNR. In other words, they show robustness over a wide range of SNR.\footnote{Although the upper bound in \eqref{equation:Channel_capacity_upper_bound} becomes tight as $\rho \to 0$, i.e., in the low-SNR regime, the proposed algorithms work well in a wide range of SNR, because both make use of the \emph{exact} channel capacity, $C(\mathbf{x},\mathbf{y})$, after solving the JCR problem.}

Finally, we investigate the effect of the size of fluid antennas on the channel capacity. According to Fig.~\ref{fig:C_vs_W}, the channel capacity increases up to a point and then reaches a peak value for large enough $W$ (channel capacity saturation). This is because the spatial-correlation parameters $\mu^{(i,j)}_{n,k} \to 0$ as $W \to \infty$. In addition, it is emphasized that significant gains in comparison with Conventional MIMO can be attained even when the fluid antenna ports are located very close to each other, usually much less than half a wavelength; note that the separation distance between the fluid antenna ports is $W\lambda /(N-1) = W\lambda /9$, which is approximately equal to $0.011\lambda \ll 0.5\lambda$ for $W=0.1$.

\vspace{-2mm}
\section{Conclusion} \label{section:Conclusion}

In this letter, we have dealt with the port selection in fluid-MIMO systems to maximize the channel capacity. Specifically, we have presented a new JCR problem and developed two optimization algorithms with different tradeoffs between performance and complexity. According to the numerical results, the proposed algorithms achieve remarkable performance.


\end{document}